\documentclass[12pt]{article}
\begin{document}
\newcommand{\beq}{\begin{equation}}
\newcommand{\eeq}{\end{equation}}
\newcommand{\beqa}{\begin{eqnarray}}
\newcommand{\eeqa}{\end{eqnarray}}
\newcommand{\beqar}{\begin{eqnarray*}}
\newcommand{\eeqar}{\end{eqnarray*}}
\newcommand{\al}{\alpha}
\newcommand{\be}{\beta}
\newcommand{\del}{\delta}
\newcommand{\D}{\Delta}
\newcommand{\eps}{\epsilon}
\newcommand{\ga}{\gamma}
\newcommand{\Ga}{\Gamma}
\newcommand{\ka}{\kappa}
\newcommand{\inn}{\!\cdot\!}
\newcommand{\h}{\eta}
\newcommand{\kk}{\varphi}
\newcommand\F{{}_3F_2}
\newcommand{\la}{\lambda}
\newcommand{\La}{\Lambda}
\newcommand{\na}{\nabla}
\newcommand{\Om}{\Omega}
\newcommand{\p}{\phi}
\newcommand{\sig}{\sigma}
\renewcommand{\t}{\theta}
\newcommand{\z}{\zeta}
\newcommand{\ssc}{\scriptscriptstyle}
\newcommand{\eg}{{\it e.g.,}\ }
\newcommand{\ie}{{\it i.e.,}\ }
\newcommand{\labell}[1]{\label{#1}} 
\newcommand{\reef}[1]{(\ref{#1})}
\newcommand\prt{\partial}
\newcommand\veps{\varepsilon}
\newcommand\ls{\ell_s}
\newcommand\cF{{\cal F}}
\newcommand\cM{{\cal M}}
\newcommand\cN{{\cal N}}
\newcommand\cH{{\cal H}}
\newcommand\cC{{\cal C}}
\newcommand\cL{{\cal L}}
\newcommand\cG{{\cal G}}
\newcommand\cI{{\cal I}}
\newcommand\cl{{\iota}}
\newcommand\cP{{\cal P}}
\newcommand\cV{{\cal V}}
\newcommand\cg{{\it g}}
\newcommand\cR{{\cal R}}
\newcommand\cB{{\cal B}}
\newcommand\cO{{\cal O}}
\newcommand\tcO{{\tilde {{\cal O}}}}
\newcommand\bz{\bar{z}}
\newcommand\bw{\bar{w}}
\newcommand\hF{\hat{F}}
\newcommand\hA{\hat{A}}
\newcommand\hT{\hat{T}}
\newcommand\htau{\hat{\tau}}
\newcommand\hD{\hat{D}}
\newcommand\hf{\hat{f}}
\newcommand\hg{\hat{g}}
\newcommand\hp{\hat{\phi}}
\newcommand\hh{\hat{h}}
\newcommand\ha{\hat{a}}
\newcommand\hQ{\hat{Q}}
\newcommand\hP{\hat{\Phi}}
\newcommand\hb{\hat{b}}
\newcommand\hc{\hat{c}}
\newcommand\hd{\hat{d}}
\newcommand\hS{\hat{S}}
\newcommand\hX{\hat{X}}
\newcommand\tR{\tilde{R}}
\newcommand\tL{\tilde{\cal L}}
\newcommand\hL{\hat{\cal L}}
\newcommand\tG{{\widetilde G}}
\newcommand\tg{{\widetilde g}}
\newcommand\tphi{{\widetilde \phi}}
\newcommand\tPhi{{\widetilde \Phi}}
\newcommand\te{{\tilde e}}
\newcommand\tk{{\tilde k}}
\newcommand\tf{{\tilde f}}
\newcommand\tF{{\widetilde F}}
\newcommand\tK{{\widetilde K}}
\newcommand\tE{{\widetilde E}}
\newcommand\tpsi{{\tilde \psi}}
\newcommand\tX{{\widetilde X}}
\newcommand\tD{{\widetilde D}}
\newcommand\tO{{\widetilde O}}
\newcommand\tS{{\tilde S}}
\newcommand\tB{{\widetilde B}}
\newcommand\tA{{\widetilde A}}
\newcommand\tT{{\widetilde T}}
\newcommand\tC{{\widetilde C}}
\newcommand\tV{{\widetilde V}}
\newcommand\thF{{\widetilde {\hat {F}}}}
\newcommand\bR{{\textbf{R}}}
\newcommand\Tr{{\rm Tr}}
\newcommand\tr{{\rm tr}}
\newcommand\STr{{\rm STr}}
\newcommand\M[2]{M^{#1}{}_{#2}}
\parskip 0.3cm

\vspace*{1cm}

\begin{center}
{\bf \Large   Generalized  Riemann curvature corrections  \\
to type II supergravity   }

\vspace*{1cm}

{  Mohammad R. Garousi\footnote{garousi@um.ac.ir} }\\
\vspace*{1cm}
{ Department of Physics, Ferdowsi University of Mashhad,\\ P.O. Box 1436, Mashhad, Iran}
\\
\vspace{2cm}

\end{center}

\begin{abstract}
\baselineskip=18pt

We observe that  the  replacement of the Riemann curvature with the generalized Riemann curvature  into the  corrections  to the type II supergravity  at order $\alpha'^3$ which are in terms of the contractions of four Riemann curvatures $R^4$,   is not fully consistent with the S-matrix elements   in the superstring theory. In particular, they produce non-zero S-matrix elements for odd number of B-field strengths which are not consistent with the string theory results. Using the consistency of the couplings with the linear T-duality as a guiding principle, we consider all T-duality invariant couplings and fix their coefficients by requiring them to be  consistent with the S-matrix elements.   The new Lagrangian density is then  equivalent to the replacement of  the generalized Riemann curvature into the expression $t_8t_8R^4$.

\end{abstract}

\setcounter{page}{0}
\setcounter{footnote}{0}
\newpage

  Many aspects of string theory can be captured at low-energy   by the Wilsonian effective action for massless fields.  The leading $\alpha'$-order terms of the effective action of type II superstring theory are given by the supergravity which contains the couplings
\beqa
S&\supset&\frac{1}{2\kappa^2}\int d^{10}x e^{-2\Phi}\sqrt{-G}\bigg[R+4(\prt\Phi)^2-\frac{1}{12}H^2\bigg]\labell{super}
\eeqa
in the NS-NS sector.   The next to the leading order couplings of the gravity are   given by the  curvature couplings  at order $\alpha'^3$  which have been found by analyzing the sphere-level four-graviton scattering amplitude in  the  superstring theory \cite{Gross:1986iv}. The result in the eight-dimensional transverse space of the light-cone formalism, is a polynomial in the linearized Riemann curvature tensors\footnote{We use only subscripts indices and the repeated indices are contracted with the inverse of metric.} 
\beqa
Y&\sim& \hat{t}_{i_1\cdots i_8}\hat{t}_{j_1\cdots j_8}R_{i_1i_2}{}_{j_1j_2}\cdots R_{i_7i_8}{}_{j_7j_8}+\cdots \labell{Y1}
\eeqa
where $\hat{t}_8$ is a tensor in eight dimensions which includes the eight-dimensional Levi-Civita tensor $\eps_{8}$ and the tensor $t_8$ that was first introduced in \cite{Schwarz:1982jn}. The contraction of $t_8$ with   four arbitrary antisymmetric matrices $M^1,\,\cdots M^4$  is defined as 
\beqa
&&t_{hkmnpqrs}M^1_{hk}M^2_{mn}M^3_{pq}M^4_{rs}=\frac{1}{\sqrt{6}}(\tr M^1M^2M^3M^4+\tr M^1M^3M^2M^4+\tr M^1M^3M^4M^2)\nonumber\\
&&\qquad\qquad\qquad -\frac{1}{4 \sqrt{6}}(\tr M^1M^2\tr M^3M^4+\tr M^1M^3\tr M^2M^4+\tr M^1M^4\tr M^2M^3)\labell{t8}
\eeqa 
 The dots in \reef{Y1} represent terms containing the Ricci and scalar curvature tensors which can not be captured by the four-graviton scattering amplitude as they are zero on-shell.  These terms  can be absorbed into the supergravity \reef{super} by appropriate field redefinition \cite{ Gross:1986iv}. 

The $t_8t_8R^4$ part of the Lagrangian \reef{Y1} 
has been  also found  in the covariant path integral formalism in \cite{Cai:1986sa}  and in the pure spinor
formalism in \cite{Policastro:2006vt}. The  Levi-Civita tensors in $\hat{t}_8\hat{t}_8$ give rise to the  covariant coupling $\eps_{10}\inn\eps_{10}R^4$ \cite{Grisaru:1986vi,Freeman:1986zh} which has its first non-zero contribution at  five graviton level \cite{Zumino:1985dp}. The presence of this term  has been dictated by  the sigma model beta function approach  \cite{Grisaru:1986vi,Freeman:1986zh}. Using the definitions of $t_8t_8$ and $\eps_{10}\inn\eps_{10}$, one finds \cite{Grisaru:1986vi,Freeman:1986zh,Myers:1987qx}
\beqa
Y
&\sim&  R_{hmnk}R_p{}_{mn}{}_qR_{hrsp}R_q{}_{rs}{}_k+\frac{1}{2}R_{hkmn}R_{pq}{}_{mn}R_{hrsp}R_q{}_{rs}{}_k+\cdots\labell{Y2}
\eeqa
where   dots represent the specific form of the  off-shell  Ricci and scalar curvature couplings which reproduce the sigma model beta function \cite{Grisaru:1986vi,Freeman:1986zh}.    It has been shown in  \cite{Garousi:2012yr} that the above Lagrangian is not consistent with the standard form of the T-duality transformations. They should be invariant under a non-standard form of T-duality transformation which receives quantum corrections. 

The $t_8t_8R^4$ part of the Lagrangian \reef{Y1}  is \cite{Gross:1986mw,Myers:1987qx,Policastro:2006vt,Policastro:2008hg}:
\beqa
\cL_1(R)&=&R_{hkmn}R_{krnp}R_{rs}{}_{qm}R_{sh}{}_{pq}+\frac{1}{2}R_{hkmn}R_{krnp}R_{rspq}R_{shqm}\labell{Y3}\\
&&  -\frac{1}{2}R_{hkmn}R_{krmn}R_{rspq}R_{shpq} -\frac{1}{4}R_{hkmn}R_{krpq}R_{rs}{}_{mn}R_{sh}{}_{pq}\nonumber\\
&& +\frac{1}{16}R_{hkmn}R_{khpq}R_{rsmn}R_{srpq}+\frac{1}{32}R_{hkmn}R_{khmn}R_{rspq}R_{srpq}\nonumber
\eeqa
 It has been shown in \cite{Myers:1987qx} that, up to field redefinition, the difference between the above two Lagrangians is the   couplings $\eps_{10}\inn\eps_{10}R^4$. The supersymmetric extension of the above Lagrangians has been studied in \cite{Paban:1998ea,Green:1998by,Peeters:2000qj}.
 
 The B-field and dilaton couplings have been added to the   Lagrangians \reef{Y1}  and \reef{Y3} by extending the linearized Riemann curvature to the generalized Riemann curvature \cite{Gross:1986mw}
 \beqa
 \bar{R}_{ab}{}^{cd}&= &R_{ab}{}^{cd}- \eta_{[a}{}^{[c}\Phi_{,b]}{}^{d]}+  e^{-\Phi/2}H_{ab}{}^{[c,d]}\labell{trans}
\eeqa
where   the bracket notation is $H_{ab}{}^{[c,d]}=\frac{1}{2}(H_{ab}{}^{c,d}-H_{ab}{}^{d,c})$, and comma denotes the partial derivative.  We will see that while the two Lagrangians \reef{Y1}  and \reef{Y3} are identical for the linearized Riemann curvature, they are not  identical for the generalized Riemann curvature. As a result, one of them should be consistent with S-matrix elements.

The   action corresponding to the Lagrangian \reef{Y3} in the Einstein frame is \cite{Gross:1986mw}
 \beqa
 S&\supset& \frac{\gamma}{\kappa^2}\int d^{10}x\sqrt{-G} e^{-3\Phi/2}\cL_1(\bar{R})\labell{S}
 \eeqa
 where $\gamma=\frac{\alpha'^3}{2^5} \z(3)$. 
To study the T-duality of the above action, one should go to the string frame in which  the   linearized $\bar{R}_{ab cd}$  becomes \cite{Garousi:2012jp}
\beqa
\bar{R}_{ab cd}=e^{-\Phi/2}\cR_{ab cd}
\eeqa
where $\cR_{ab cd}$ is the following expression 
\beqa
\cR_{ab cd}&=&R_{ab cd}+H_{ab [c,d]}\labell{RH2}
\eeqa
It has the symmetries $\cR_{bacd}=-\cR_{abcd}$ and $\cR_{abdc}=-\cR_{abcd}$.  The action  \reef{S} in the string frame then becomes
\beqa
 S&\supset& \frac{\gamma}{\kappa^2}\int d^{10}x\sqrt{-G} e^{-2\Phi}\cL_1(\cR)\labell{S1}
 \eeqa 
 The dilaton appears only as the overall   factor  $e^{-2 \Phi}\sqrt{-G}$  which is invariant under standard   T-duality. It has been shown in \cite{Garousi:2012jp} that the  Lagrangian $\cL_1(\cR)$ is  also invariant under  the  standard  linear T-duality transformations.
 
The Lagrangian $\cL_1(\cR)$ contains the couplings $R^4$, $H^4$ and $H^2R^2$ which are exactly reproduced by string theory S-matrix elements \cite{Gross:1986mw}. However, it contains also the couplings $R^3H$ and $RH^3$ which are not reproduced in string theory. One can easily verify that the supergravity does not produce scattering amplitude of odd number of B-field strengths.  Therefore, the string theory S-matrix element which reproduces the supergravity results at the leading order of $\alpha'$,  is zero for odd number of $H$. 

To check that the Lagrangian $\cL_1(\cR)$ produces the non-zero couplings for odd number of $H$, one should first replace the generalized Riemann curvature \reef{RH2} in \reef{S1}. It produces, for example,  24  couplings between three $H$ and one Riemann curvature, \ie
\beqa
S_{HHHR}= \frac{\gamma}{\kappa^2}\int d^{10}x\sqrt{-G} e^{-2\Phi}\bigg[-\frac{1}{4} H_{s h[p,q]} H_{k r[p,q]} H_{r s[m,n]} R_{h k m n}+\cdots\bigg]
 \eeqa
where dotes represent the other 23 terms. To verify that the above couplings do not simplify to zero, 
 one may write the linearized Riemann curvature and the field strength $H$ as 
\beqa
R_{\mu \nu \alpha \beta}&=& \kappa(h_{\mu \beta,\nu \alpha}+h_{\nu \alpha,\mu \beta}-h_{\mu  \alpha,\nu \beta}-h_{\nu \beta,\mu \alpha})\nonumber\\
H_{\mu \nu [\alpha, \beta]}&=& \kappa(b_{\mu \beta,\nu \alpha}+b_{\nu \alpha,\mu \beta}-b_{\mu  \alpha,\nu \beta}-b_{\nu \beta,\mu \alpha})\labell{RH}
\eeqa
where as usual  the comma represents partial differentiation. The graviton $h_{\mu\nu}$ and the antisymmetric tensor $b_{\mu\nu}$ may be written as
\beqa
h_{\mu\nu}&=&\frac{1}{2}(\psi_{\mu}\zeta_{\nu}+\psi_{\nu}\zeta_{\mu})\nonumber\\
b_{\mu\nu}&=&\frac{1}{2}(\psi_{\mu}\zeta_{\nu}-\psi_{\nu}\zeta_{\mu})
\eeqa
where $\psi$ and $\zeta $ are two vector fields. Then one may transform the couplings to the momentum space. To this end, one should label the antisymmetric fields by 1,2,3 and the graviton by 4. Then one should add the 6 permutations of the antisymmetric fields. Performing all these steps, one finds that the result is not zero. Even if one uses the on-shell relations $k_i\inn k_i=0$ and $k_i\inn \eps_i=0$ where $\eps_i$ is the polarization of the $i$-th particle, the couplings still do not vanish. Doing the same steps for the $HR^3$ couplings, one again finds non-zero couplings. In fact the couplings of odd number of $H$ resulting from the terms in the second line of \reef{Y3} remain non-zero at the linearized level. 

Since the Lagrangian $\cL_1(\cR)$ is not fully consistent with the string theory S-matrix elements, one expects there must be another Lagrangian with the following properties:

1-It should produce no couplings $H^4$, $R^4$ or $R^2H^2$.

2-It should  produce the   couplings $R^3H$ and $RH^3$ which cancel the corresponding couplings in $\cL_1(\cR)$.

3-It should be consistent with the standard T-duality.

One may consider all possible contractions  of four generalized Riemann curvatures, and may choose unknown coefficient for each of them. Then one may find the coefficients by forcing them to satisfy the above constraints. 

To impose the T-duality constraint, we note that under linear T-duality the Riemann curvature with two Killing indices transforms as \cite{Garousi:2012yr}
\beqa
R_{\mu y \nu y}&\rightarrow &-R_{\mu y\nu y} 
\eeqa 
where $y$ is the killing index. So under the dimensional reduction on a circle, the couplings with structure $RR_{yy}R_{yy}R_{yy}$ where $R_{yy}$ is the Riemann curvature with two Killing indices, are not consistent with the linear T-duality. To avoid such couplings we consider the contractions of the generalized Riemann curvature in which the first two indices of the curvatures contract among themselves, and the second two indices contract among themselves, as the couplings in \reef{Y3}. Using the symmetries of the curvature $\cR_{abcd}$, one finds there are eight independent such couplings. Considering them with unknown coefficients, and constraining them to satisfy the conditions 1 and 2, one finds the following couplings:
\beqa 
\cL_2(\cR)&=&  -\frac{1}{8} \cR_{h k m n} \cR_{k p r s} \cR_{h q r s} \cR_{p q m n}+\frac{1}{8} \cR_{h k m n} \cR_{h k r s} \cR_{p q n r} \cR_{p q m s}\nonumber\\&&
-\frac{1}{4} \cR_{h k m n} \cR_{k p m n} \cR_{h q r s} \cR_{p q r s}-\frac{1}{4} \cR_{h k m n} \cR_{h k n s} \cR_{p q m r} \cR_{p q r s}\labell{Y4}
\eeqa
In fact $\cL_2(R)=0$  for Riemann curvature, however, it is not an identity any more  for the generalized Riemann curvature. The Lagrangian $\cL_1(\cR)+\cL_2(\cR)$ now has only couplings $H^4$, $R^4$ and $R^2H^2$. It has been shown in \cite{Garousi:2012jp} that such couplings at four-field level are consistent with the linear T-duality. 

Since there are eight independent couplings in which the first two indices of the Riemann curvatures contract among themselves, two of the above couplings must have the same structure as the terms in \reef{Y3}. The first and the third terms in \reef{Y4} have the same structure as the terms in the second line of \reef{Y3} but their coefficients are different. Adding the Lagrangian $\cL_1$ to $\cL_2$, one finds that  all independent contraction of the four generalized Riemann curvatures which are consistent with the linear T-duality have non-zero coefficients.  Therefore, the action in the string frame which is consistent with the linear T-duality and is fully consistent  with the four-point functions of string theory, is 
\beqa
 S&\supset& \frac{\gamma}{\kappa^2}\int d^{10}x\sqrt{-G} e^{-2\Phi} \cL(\cR) \labell{S2}
 \eeqa  
 where the Lagrangian density has the following eight independent terms:
 \beqa
 \cL(\cR)&=& \cR_{h k m n} \cR_{k r n p}\cR_{r s m q}R_{h s p q}+\frac{1}{2} \cR_{h k m n} \cR_{k r n p}\cR_{r s p q} \cR_{h s m q}\nonumber\\&&-\frac{1}{4} \cR_{h k m n} \cR_{h k n s} \cR_{p q m r} \cR_{p q r s}+\frac{1}{8} \cR_{h k m n} \cR_{h k r s} \cR_{p q n r} \cR_{p q m s}\nonumber\\&&+\frac{1}{4} \cR_{h k m n} \cR_{k r m n}\cR_{r s p q} \cR_{h s p q}+\frac{1}{8} \cR_{h k m n} \cR_{k r p q}\cR_{r s m n} \cR_{h s p q}\nonumber\\&&+\frac{1}{16} \cR_{h k m n} \cR_{h k p q} \cR_{r s m n} \cR_{r s p q}+\frac{1}{32} \cR_{h k m n}\cR_{h k m n} \cR_{r s p q}\cR_{r s p q}\labell{L}
 \eeqa
Note that the couplings in the first and the last lines above are the same couplings in \reef{Y3}. 

Now let us compare the above Lagrangian with the replacement of the generalized Riemann curvature \reef{RH2} into $t_8t_8R^4$.  
Using the definition of the tensor $t_8$ in \reef{t8}, and using the fact that the generalized Riemann curvature $\cR_{abcd}$ has the same symmetries of the Riemann curvature expect the symmetry under $(ab)\leftrightarrow (cd)$, one finds after some algebra
\beqa
t_8t_8\cR^4&=& \cL(\cR)\labell{final}
\eeqa
Therefore, the replacement \reef{RH2} in the Lagrangian $t_8t_8R^4$ is consistent with the S-matrix elements of four NS-NS vertex operators and with the linear T-duality. 

The Lagrangian \reef{L} may be  extended to nonlinear order by replacing the linearized $\cR$  with the nonlinear generalized Riemann curvature
\beqa
\cR_{abcd}&\rightarrow&R_{abcd}+H_{ab[c;d]}+\frac{1}{2}H_{ae[c}H_{be|d]}
\eeqa
One can easily verify that the last term has the symmetries of the Riemann curvature, so the above replacement in \reef{L} does not produce odd number of $H$, as expected. It would be interesting to compare the couplings $H^2R^3$ and $H^4R$ resulting from the above replacement,     with the contact terms of the corresponding sphere-level S-matrix element in string theory. At the one-loop level of type IIA theory, it has been shown in \cite{Liu:2013dna} that the above replacement in $t_8t_8R^4$ and $B_2\wedge X_8$ are consistent with S-matrix calculation and with the T-duality, however, this replacement in $\eps_{10}\inn\eps_{10}R^4$ is not  consistent with the S-matrix calculation. 

  {\bf Acknowledgments}:    This work is supported by Ferdowsi University of Mashhad under grant 2/20625.


\end{document}